\documentclass[%
reprint,
superscriptaddress,
 amsmath,amssymb,
 aps,
 pra,
]{revtex4-1}

\usepackage{multibib}

\usepackage{graphicx}
\usepackage{dcolumn}
\usepackage{bm}

\usepackage[utf8]{inputenc} 

\usepackage{multirow}
\usepackage{import}
\usepackage[warn]{mathtext}
\usepackage{amsfonts}
\usepackage{amsmath}
\usepackage{amssymb}
\usepackage{braket}
\usepackage{bbold}
\usepackage{textcomp} 
\usepackage{indentfirst} 
\usepackage{amsmath} 
\usepackage{siunitx}
\usepackage{graphicx}
\usepackage [T2A]{fontenc}
\usepackage [utf8]{inputenc}
\usepackage[english]{babel}

\DeclareGraphicsExtensions{.pdf,.png,.jpg}

\usepackage{hyperref}

\usepackage{algpseudocode}

\usepackage[nottoc]{tocbibind}

\usepackage{xcolor}
\hypersetup{
    colorlinks,
    linkcolor={red!50!black},
    citecolor={blue!50!black},
    urlcolor={blue!80!black}
}



\begin{document}

    \title{Effect of Etching Methods on Dielectric Losses in Transmons}

    \author{Chudakova~T.~A.}
    \affiliation{National University of Science and Technology MISIS, Moscow, 119049 Russia}
    \affiliation{Russian Quantum Center, Skolkovo, Moscow, 121205 Russia}
    \affiliation{Moscow Institute of Physics and Technology, Dolgoprudnyi, Moscow region, 141701 Russia}

    \author{Mazhorin~G.~S.}
    \affiliation{National University of Science and Technology MISIS, Moscow, 119049 Russia}
    \affiliation{Russian Quantum Center, Skolkovo, Moscow, 121205 Russia}
    \affiliation{Moscow Institute of Physics and Technology, Dolgoprudnyi, Moscow region, 141701 Russia}
    
    \author{Trofimov~I.~V.}
    \affiliation{Institute of Nanotechnology of Microelectronics, Russian Academy of Sciences, Moscow, 119991 Russia}

    \author{Rudenko~N.~Yu.}
    \affiliation{National University of Science and Technology MISIS, Moscow, 119049 Russia}
    
    \author{Mumlyakov~A.~M.}
    \affiliation{Institute of Nanotechnology of Microelectronics, Russian Academy of Sciences, Moscow, 119991 Russia}

    \author{Kazmina~A.~S.}
    \affiliation{National University of Science and Technology MISIS, Moscow, 119049 Russia}
    \affiliation{Russian Quantum Center, Skolkovo, Moscow, 121205 Russia}
    \affiliation{Moscow Institute of Physics and Technology, Dolgoprudnyi, Moscow region, 141701 Russia}

    \author{Egorova~E.~Yu.}
    \affiliation{National University of Science and Technology MISIS, Moscow, 119049 Russia}
    \affiliation{Russian Quantum Center, Skolkovo, Moscow, 121205 Russia}
    \affiliation{Moscow Institute of Physics and Technology, Dolgoprudnyi, Moscow region, 141701 Russia}

    \author{Gladilovich~P.~A.}
    \affiliation{National University of Science and Technology MISIS, Moscow, 119049 Russia}

    \author{Chichkov~M.~V.}
    \affiliation{National University of Science and Technology MISIS, Moscow, 119049 Russia}

    \author{Maleeva~N.~A.}
    \affiliation{National University of Science and Technology MISIS, Moscow, 119049 Russia}
    
    \author{Tarkhov~M.~A.}
    \affiliation{Institute of Nanotechnology of Microelectronics, Russian Academy of Sciences, Moscow, 119991 Russia}
     
    \author{ Chichkov~V.~I.}
    \affiliation{National University of Science and Technology MISIS, Moscow, 119049 Russia}
    
\date{\today}

 \begin{abstract}
Superconducting qubits are considered as a promising platform for implementing a fault tolerant quantum
computing. However, surface defects of superconductors and the substrate leading to qubit state decoherence
and fluctuations in qubit parameters constitute a significant problem. The amount and type of defects depend
both on the chip materials and fabrication procedure. In this work, transmons produced by two different
methods of aluminum etching: wet etching in a solution of weak acids and dry etching using a chlorine-based
plasma are experimentally studied. The relaxation and coherence times for dry-etched qubits are more than
twice as long as those for wet-etched ones. Additionally, the analysis of time fluctuations of qubit frequencies
and relaxation times, which is an effective method to identify the dominant dielectric loss mechanisms,
indicates a significantly lower impact of two-level systems in the dry-etched qubits compared to the wetetched ones.
\end{abstract}
	
\maketitle

\section{INTRODUCTION}

Superconducting qubits are one of the most promising platforms for implementing quantum computing. To date, qubits with millisecond coherence times \cite{Somoroff_2023},  high-fidelity single-qubit  \cite{li2023error} and two-qubit gates \cite{Moskalenko2022, ding2023highfidelity}  have been featured. Furthermore, a quantum supremacy has already been demonstrated on superconducting quantum processors \cite{arute2019quantum, PhysRevLett.89.137901}, as well as error
correction algorithms \cite{google2023suppressing, Krinner_2022},  and solutions of the first
quantum chemistry problems \cite{QChemistry1,QChemistry2}. However, limited coherence of the qubits, mostly due to the presence of defects, remains one of the main obstacle for
the superconducting systems on the way to the fault
tolerant quantum computing  \cite{Mul1, Relax_evidance, McDermott, Béjanin, Pappas}.

Superconducting qubits fabricated on a dielectric
substrate are coupled to defects by an electric field,
which induce the relaxation of qubits \cite{Klimov}. 
The concentration and type of defects depend on both the
superconductor and the substrate materials, as well as
on the treatment of their surfaces  \cite{Niepce, Burnett, Müller2, bib2}.  A model defect is a two-level system (TLS), where states are separated by an energy of about $10^{-5}$ eV. Such TLS transitioning from one state to another can absorb
qubit energy leading to the relaxation of the qubit state
or to a change in the qubit parameters resulting in the
loss of coherence \cite{Al1, Müller3}. In the limit of many nonresonant non-interacting defects weakly coupled to
the qubit, qubit state relaxation can be well characterized by the dielectric loss tangent \cite{Dielec_loss, Smirnov_2024}.

Two-level systems with frequencies close to the
qubit one must be considered separately  \cite{Müller4}. Strongly coupled resonant defects can interact coherently with
the qubit, leading to phenomena such as beats in Rabi
or Ramsey oscillations \cite{Direct_TLS}.
On the other hand,
weakly coupled resonant defects cause relaxation of
the qubit and could mediate the interaction with other
low-frequency TLSs. Such interaction can cause 1/f
noise leading to dephasing, fluctuations, and jumps in
the qubit coherent characteristics and its frequency
over time \cite{Shnirman, Dutta, Lisenfeld,  Resonator_flikker, Klimov, 1_f}.

One of the ways to reduce the influence of defects
is to change qubit topology in order to reduce the
interaction with noise sources \cite{shape_opt}.
Another way is to
use signals that mitigate the interaction between the
qubit and specific defects, for example detuning the
frequencies of qubits from parasitic TLSs or control
the frequencies of these TLSs using an electric field \cite{El_TLS}. The above methods to mitigate the impact of
defects require additional hardware and complicate
device management, making scaling even more challenging problem. Furthermore, while performing twoqubit operations, auxiliary levels \cite{Gate_fluxonium} or frequency
tuning of processor elements \cite{Gate_fluxonium2} are often used, reducing the efficiency of these methods in real applications.

The described techniques allow the reduction of
dielectric losses for a given number of defects. However, a more effective approach is to decrease the number of defects by employing advanced qubit fabrication
techniques \cite{bib8, bib9}. In this work, we experimentally
study the decoherence of transmons, fabricated by two
different techniques of thin aluminum films etching:
dry plasma-chemistry and wet etching.

\section{TOPOLOGY AND FABRICATION
OF SAMPLES}

\begin{figure}[!h]
    \centering
    \includegraphics[width=0.48\textwidth]{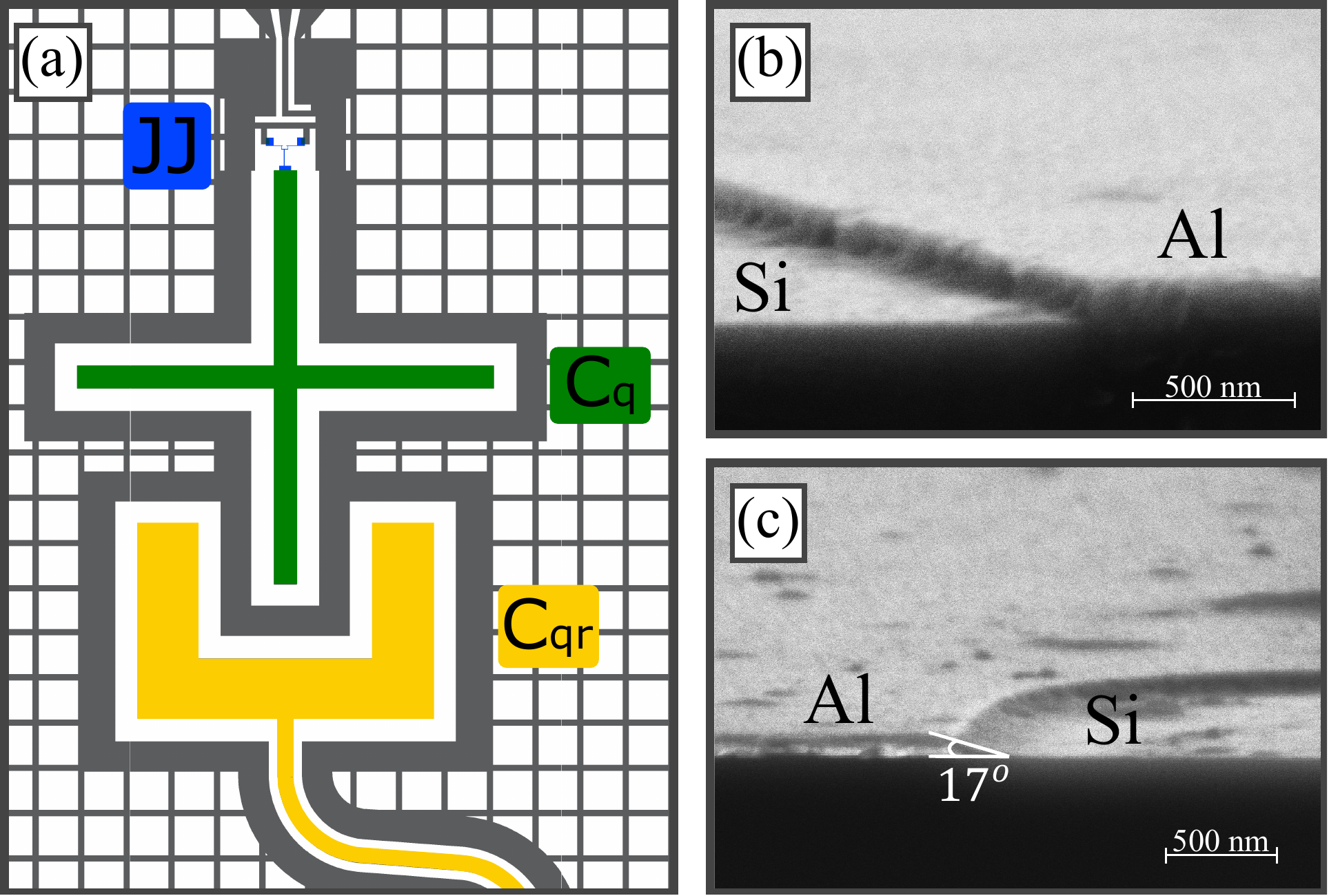}
    \caption{ 
    (Color online) (a) Topology of the cross-shaped
transmon. The Josephson junction SQUID is shown in
blue, the shunt capacitor is green, and a fragment of the
individual readout resonator is yellow. (b, c) Etching profiles of the aluminum film obtained by (b) dry plasma
etching and (c) wet etching methods.}
    \label{fig:topology}
\end{figure}

In order to study the impact of the etching process
on the magnitude and nature of dielectric losses, two
samples each containing five cross-shaped transmons
(Xmons) were fabricated. The topology of a single
qubit is presented in  Fig. \ref{fig:topology} (a).The qubit consists of a two-contact asymmetric SQUID, shown in blue,
shunted by a large cross-shaped capacitor, shown in
green. Measurements were performed at zero magnetic flux applied to the SQUID, i.e. at the point of
the lowest Xmon sensitivity to flux noise. The qubit is
capacitively coupled to an individual readout resonator, shown in yellow, which is inductively coupled to
the readout line. The readout resonator is also used for
the qubit excitation. The above design was chosen due
to the simple fabrication resulting in a high reproducibility and most important for this study, the relaxation
and coherence times of such qubits are typically limited by dielectric losses rather than by relaxation to
other circuit elements.

We used high-resistivity silicon ($\rho>10000~$k$\Omega$$\cdot$cm). 
 as
the substrate employing standard cleaning methods
involving deionized water and organic solvents for the
surface preparation. Prior to aluminum deposition,
the natural oxide layer on the silicon surface was
removed using an HF buffer solution (BOE 7:1 VLSI) to ensure cleanliness. 
A 100-nm-thick aluminum film
was deposited by electron beam evaporation machine
with residual pressure less than 10$^{-8}$ mbar.  After the
aluminum deposition, the substrate was cut into two
samples A and B. The S1805 photoresist was utilized to
create the basic structural pattern and the pattern was
transferred using direct laser photolithography.

The next step was the etching of the aluminum
film. For sample A, we used a dry plasma chemical
method: reactive ion etching was performed in a reactor with capacitively and inductively coupled plasma (ICP-RIE)containing gas mixture of ($BCl_3$), (HBr), and ($Cl_2$) in a 1:2:2 ratio. 
These gases play a crucial role in
the formation of volatile chemical compounds with
aluminum and aluminum oxide. During the etching
process, gaseous chlorine was ionized into $Cl^-$ ions,
leading to the formation of the volatile $AlCl_3$ compound. Aluminum oxide formed during the etching
was removed using $BCl_3$, providing $B_2O_3$ and volatile  $AlCl_3$. In order to achieve higher selectivity with regard
to the photoresist and create an anisotropic profile,
hydrogen bromide was added as an additional gas. The
table temperature during the etching was 50$\si{\celsius}$, the
pressure in the chamber was 2  mTorr, the ICP power
was 600 W and the RF power was 15 W.The process
lasted for 1 min. The aluminum etching rate was
102 nm/min and the selectivity towards photoresist
was 2.25. Sample B was etched in a solution of $H_3PO_4 : HNO_3 : CH_3COOH : H_2O = 73 \% : 3.1 \% : 3.3 \%: 20.6.\% $  (TechniEtch Al80) at 50$\si{\celsius}$  for 20 sec.
Fig. \ref{fig:topology} demonstrates the etching profiles of the aluminum film by dry (b) and wet (c) methods. As a result
of dry plasma chemical etching, the profile has a sharp
step with characteristic roughnesses \cite{Rodionov},while wet
etching produced a smooth, sloping boundary that
ends in an angle of approximately 17$\si{\celsius}$. Additionally,
the quality of surfaces after dry etching is better than
that of wet etching.

Further technological steps were the same for both
samples. After the mask was removed in N-methyl-2-
pyrrolidone (NMP) at 80$\si{\celsius}$, two layers of
PMMA/MMA photoresists were applied. 
Then, using electron beam lithography and the bridge-free fabrication~\cite{Lecocq2011}, the Josephson junctions Al/AlO$_x$/Al were
formed. For galvanic contact between the junctions
and the base layer, two aluminum electrodes were
deposited with intermediate oxidation. This aluminum bandages were formed by the deposition of aluminum through a photoresistive mask on the surface
of the base structure cleaned with Ar ions. After the
lift-off in NMP at 80$\si{\celsius}$, both samples were cut into individual chips on a disk cutting machine.

The target parameters of the Xmons were a capacitance of 67~fF, and Josephson energies of 9.6 and 5.8~GHz. These parameters were measured using spectroscopy methods. The deviation from the designed values was within $10 \% $ and the scatter relative to the average was within $2 \% $ for both dry- and wetetched samples.

\section{ EXPERIMENTAL CHARACTERIZATION
OF THE QUBITS}

The purpose of the reported experiments was to
measure the relaxation and coherence times of qubits
and analyze noise sources that affect the coherence of
qubits. One way of identifying these noise sources is to
measure fluctuations of the qubit coherent characteristics over time. 

In order to measure these characteristics, the calibration of the single-qubit control $\pi$- and $\pi/2$-pulses \cite{devoret2013superconducting}used in the experiment is needed. The duration of
these pulses is determined by measuring Rabi oscillations, which are oscillations in the qubit population
depending on the pulse duration \cite{Krantz_2019}. Relaxation, $T_1$ and coherence,  $T_2$, times measured using Rabi and Ramsey protocols are summarized in Table \ref{table:Times}.

\begin{table}[tbh]
\caption{\label{table:Times}  Relaxation, $T_1$ , and coherence, $T_2$ , times of samples produced using dry plasma chemical (sample A) and wet (sample B) etching methods. }
\vspace{2mm}
\centerline{
\begin{tabular}{m{1.2cm}|m{1.6cm} m{1.6cm}|m{1.6cm} m{1.6cm}}
\hline
\multirow{2}{3em}{Qubit}
& \multicolumn{2}{c|}{Sample А: dry} & \multicolumn{2}{c}{Sample В: wet} \\
& $T_1$, $\mu s$ & $T_2$, $\mu s$ & $T_1$, $\mu s$ & $T_2$, $\mu s$\\[0.2ex]
\hline\hline
1      &  56   &  45   &   20  &  14 \\
2      &  50   &  37   &   15  &  11\\
3      &  54   &  37   &   21  &  20  \\
4      &  64   &  17  &  22  &  6  \\
5      &  75   &  32   &   20  &  18  \\
\hline
\end{tabular}}
\end{table}

The relaxation and coherence times of the wetetched sample B qubits are smaller than those for the
dry-etched qubits by a factor of 2.5 -- 3 on average. This
may be caused by the strong electrostatic interaction
between the qubit and defects on the surface and edge
of the superconductor formed during etching.

\begin{figure}[!h]
    \centering
    \includegraphics[width=0.48\textwidth]{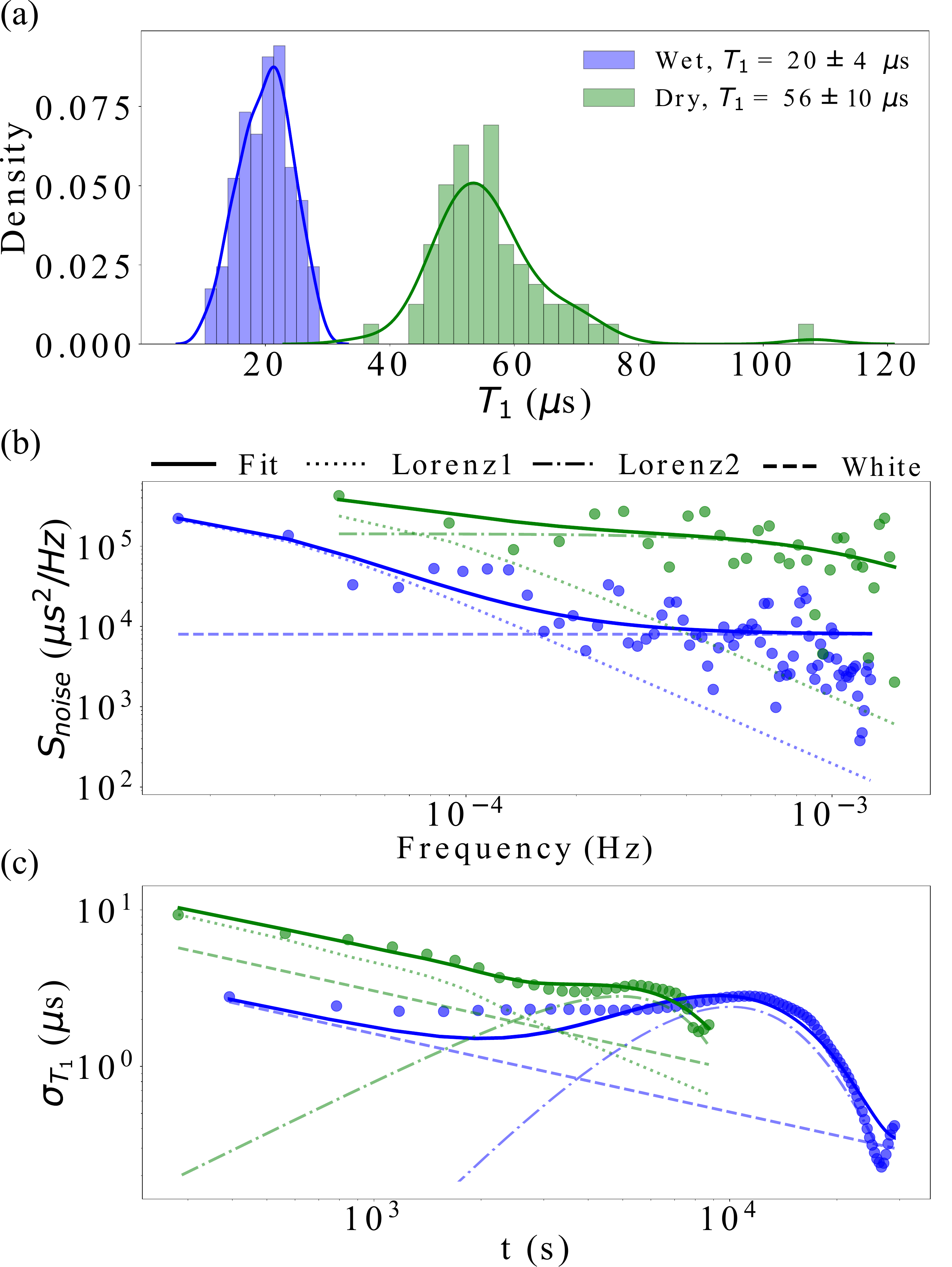}
    \caption{ (Color online) Results of repeated measurements of the relaxation times T1 for samples A and B are given in
green and blue, respectively. (a) Relaxation time histograms. (b) Welch spectral analysis of the fluctuations. The
different line styles represent the experimental data, the
individual contributions from white and Lorentzian noise,
and their sum. (c) Allan deviation of the fluctuations in the
relaxation times; the line styles are the same as in panel (b).}
    \label{fig:T1}
\end{figure}

In order to better understand the reasons for the
significantly improved coherence of the dry-etched
sample A, further analysis was performed. We
repeated measurements of the relaxation time, coherence time, and frequency for both samples, but with
different measurement durations. For the sample A,
we measured the relaxation time for 5.5 min and the
coherence time for 4 min, while for the sample B we
measured relaxation time for 6.5 min and coherence
time for 4 min. Relaxation time histograms for the first
qubits in both samples are shown in Fig. \ref{fig:T1} (a).  In order to determine the sources of noise that cause fluctuations in the qubit relaxation time, we calculated the
Allan deviation and performed a Welch spectral analysis  ~\cite{bib13} , as shown in Figs. \ref{fig:T1}~(b) and (c).

The Allan deviation, by definition, is an average of
the squared difference between averaged consecutive
samples with a given duration; in this case,

\begin{equation}
\begin{gathered}
\label{eq24}
\sigma_{T_1}^2 = \frac{1}{2} \left\langle{\left( {\langle T_1 \rangle}_{\tau}^{n+1} - {\langle T_1 \rangle}_{\tau}^n \right) ^2}\right\rangle
\end{gathered}
\end{equation}

The presence of peaks in the Allan deviation indicates the existence of Lorentzian noise in the system.
Using combined data from deviations and spectral
analysis, we determined the contributions of the white
noise and Lorentzian noise to fluctuations in relaxation times. In Figs. \ref{fig:T1}~(b) and (c), the solid line represents the simulation result, which shows good agreement with experimental data for both wet etched qubits (blue line) and dry etched qubits (green line).
The Lorentzian curve in the fluctuation spectrum typically corresponds to the presence of a resonance in
the system. This indicates coupling to a system that has
a specific switching frequency. Such systems can include quasiparticle excitations or TLS. We know
that the characteristic frequencies for quasiparticle
recombination at low temperatures are around 1 kHz ~\cite{bib6}, while quasiparticle tunneling in Josephson junctions ranges from 0.1 to 10 kHz~\cite{bib7}.In the experiment with the wet etched qubit, the Lorentzian switching
frequency was 40 $\mu$Hz.The deviation of the dryetched qubit had two distinguishable Lorentzian peaks with switching frequencies of 75 and 800 $\mu$Hz. These
observed switching frequencies were significantly
lower than typical values for quasiparticles, but they
are typical of low-frequency fluctuators~\cite{bib8, bib9}.  The interaction of these systems through a high-frequency
defect associated with the qubit was the most likely
source of the Lorentzian noise observed in the experiment.

\begin{figure}[!h]
    \centering
    \includegraphics[width=0.48\textwidth]{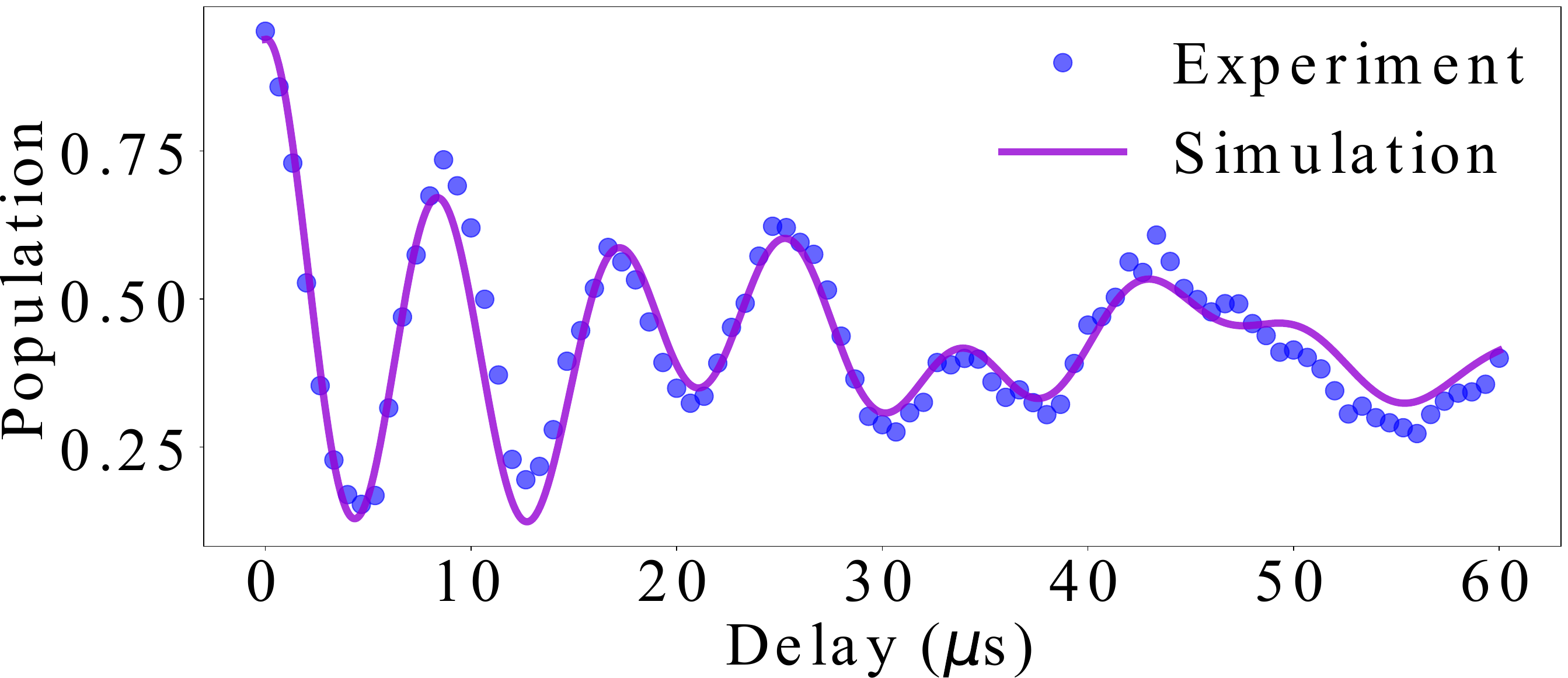}   
    \caption{  (Color online) Experimental points for Ramsey
oscillations of the first qubit in sample B in comparison
with the numerical simulation of the qubit–defect system.
    }
    \label{fig:Ramsey_single}
\end{figure}

Highly coherent TLSs that are strongly connected
to a qubit can cause fluctuations in relaxation time and
participate in the coherent dynamics of the system. Fig.~\ref{fig:Ramsey_single}  shows the results of Ramsey oscillation measurements on the first qubit in a sample fabricated using wet etching. The experimental points are clearly separated from the curve of exponentially decaying
sinusoidal oscillations, indicating the presence of
“beats.” The explanation for these beats is a coherent
exchange of population with a high-frequency,
strongly coupled TLS. To confirm this explanation,
the strength of coupling between the defect and qubit
was measured to be 26 kHz and the detuning between
the qubit frequency and the defect frequency was
found to be 54 kHz. Modeling of the qubit-defect system (shown in Fig. ~\ref{fig:Ramsey_single} as the solid line) provides a good
fit to the experimental data. The electric dipole coupling between the qubit and the defect, obtained
through numerical modeling, allows us to estimate the
distance from the Josephson junction to the defect~\cite{dip_dist} as 10 $\mu$m or less.

\begin{figure}[!h]
    \centering
    \includegraphics[width=0.45\textwidth]{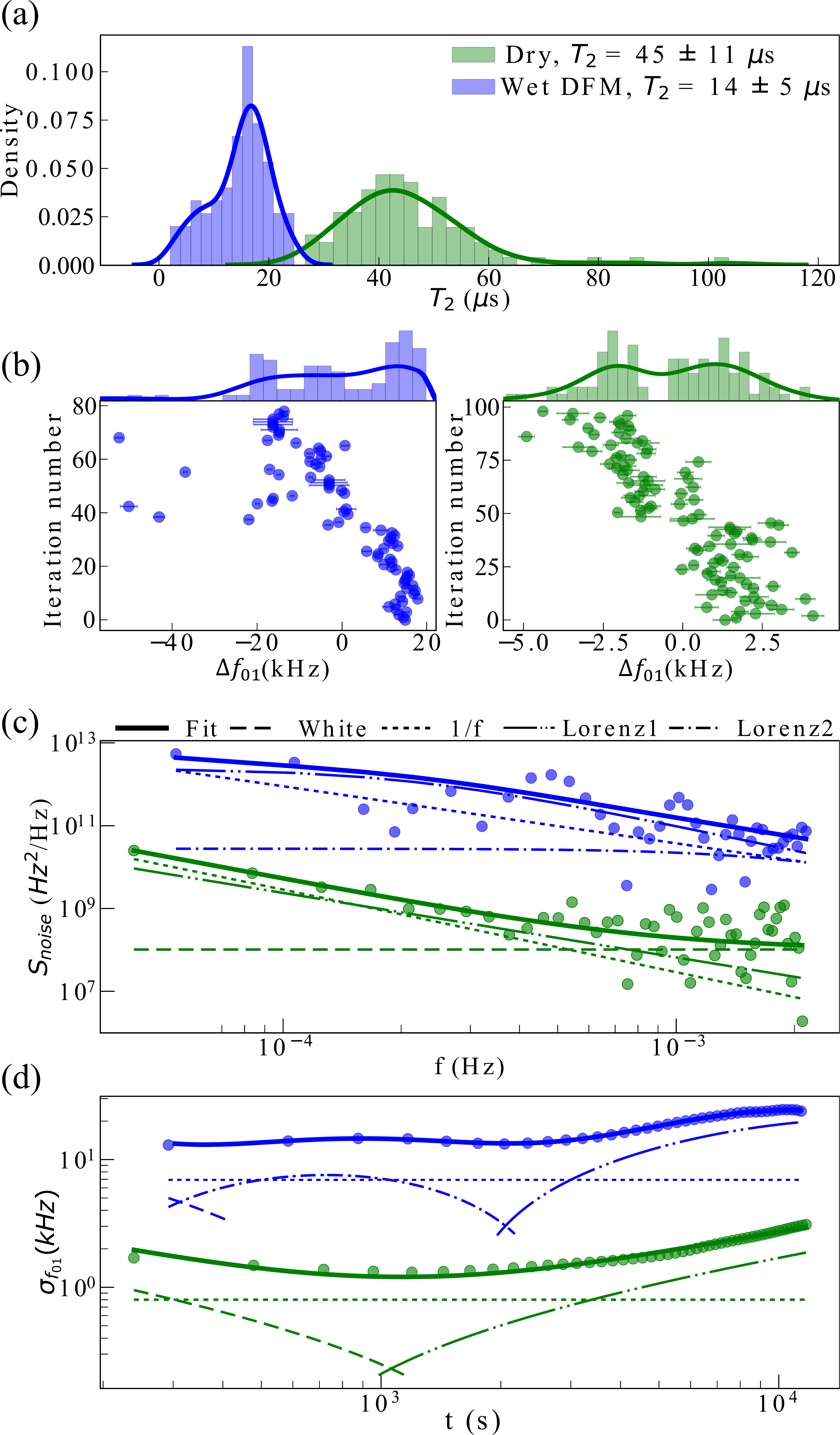}
    \caption{ (Color online) Results of the Ramsey experiments
for samples A and B are given in green and blue, respectively. (a) Coherence time histogram. (b) Fluctuations in
the qubit frequency. The qubit frequency in the wet-etched
sample drifts at a rate of 6 kHz/h, while the qubit frequency in the dry-etched sample exhibits switching in the
5 kHz range. (c) Welch spectral analysis of the frequency
fluctuations. The different lifestyles represent experimental data, individual contributions from white, 1/f,
and Lorentzian noise, and their sum. (d) Allan deviation
of frequency fluctuations; the line styles are the same as in
panel (c).
    }
    \label{fig:Ramsey_stat}
\end{figure}

The results of Ramsey oscillation measurements
for the first qubits of each sample are presented in
Fig.~\ref{fig:Ramsey_stat}. 
The contributions from white, 1/f, and Lorentzian noise were determined from the cumulative data of the Allan deviation shown in Fig. ~\ref{fig:Ramsey_stat}(d)  and from the Welch noise spectrum presented in Fig.~\ref{fig:Ramsey_stat}(c). The
switching frequencies for Lorentzian curves are 106 $\mu$Hz and 2 mHz for the wet-etched sample and 6 $\mu$Hz for the dry etched sample.

As mentioned above, the most likely sources of the noise may be the low-frequency fluctuators. These fluctuators are manifested not only as Lorentzian noise. We observed a frequency drift at a rate of 6 kHz/h and high levels of 1/f noise for the wet-etched sample B. This may be a result of interaction with a large number of low-frequency fluctuators \cite{1_f}. We observed jumps in the form of telegraph noise between two pronounced frequencies in the dry-etched sample A. This can be caused by a single low-frequency
fluctuator switching \cite{Schlör}.

\section{MEASUREMENTS OF THE NOISE SPECTRAL DENSITY}

Measurements of the time fluctuations in the
coherent characteristics can be useful for analyzing
noise sources. However, they only allow to study lowfrequency noise (characteristic frequencies below
10 mHz), which are not averaged over the duration of
the measurement. In order to investigate the effect of
high-frequency noise, a spin-locking protocol was
employed \cite{Spin_loc}, This protocol allows the measurement
of the noise spectral density over a frequency range
from kHz to 10 MHz.

\begin{figure}[!h]
    \centering
    \includegraphics[width=0.48\textwidth]{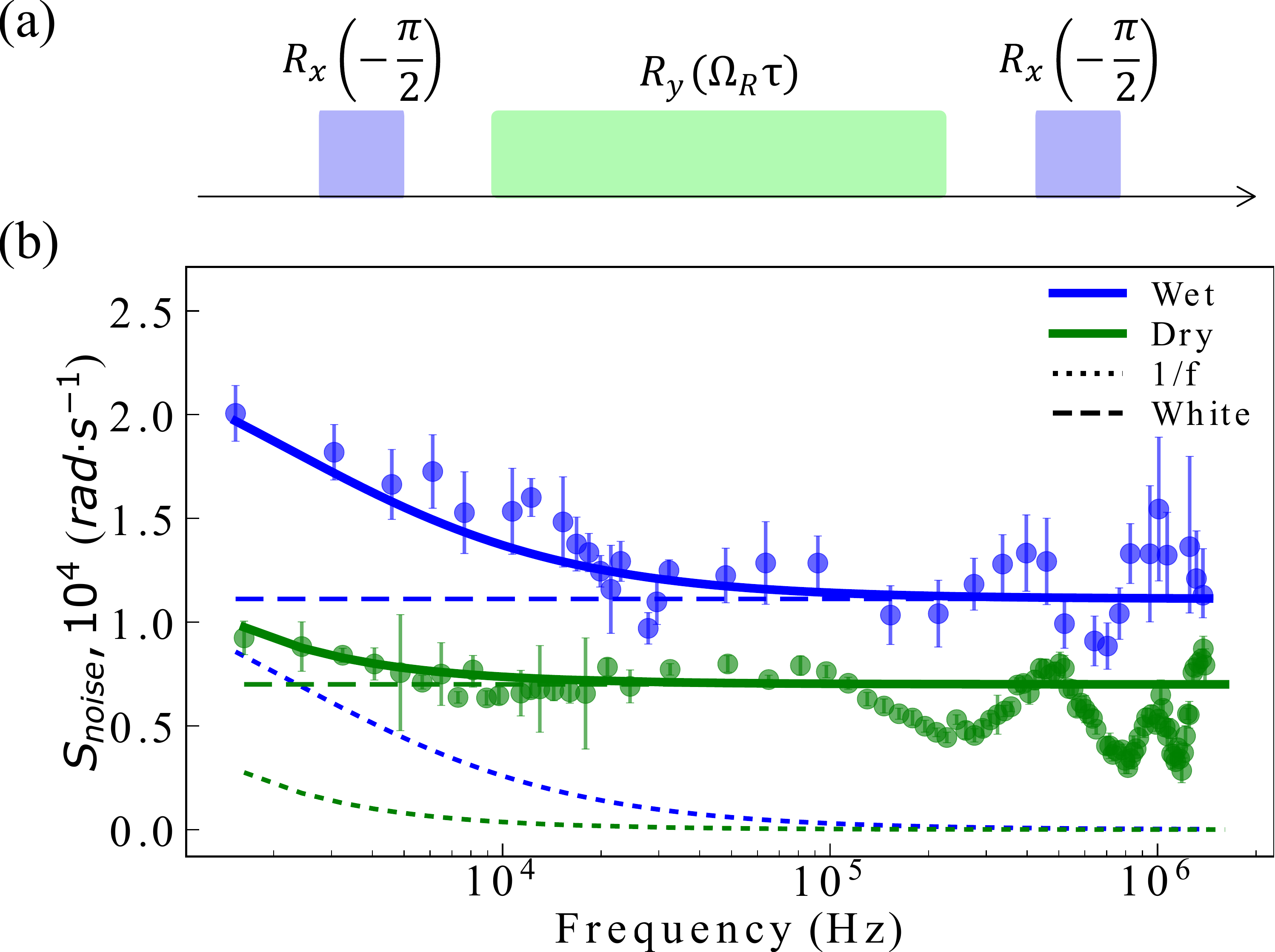}
    \caption{(Color online) Noise spectral density. (a) Spinlocking pulse sequence used to obtain the noise spectral density. (b) Measured noise spectral densities for the wetand dry-etched samples are given in green and blue, respectively. The different line styles correspond to experimental data, the individual contributions from white and 1/f noise, and their sum.}
    \label{fig:spin_lock}
\end{figure}

Fig.~\ref{fig:spin_lock} (a) illustrates the pulse sequence for the spin-locking protocol. Two $R_x(\pi/2)$ pulses are
applied to the qubit, with a rectangular  $R_y(\Omega_R\tau)$ pulse in between, followed by the qubit state measuring
depending on the duration $\tau$ and amplitude of the  $R_y(\Omega_R\tau)$ pulse.
The phases of the two $\pi/2$ pulses are
the same, and shifted by $\pi/2$ from the $R_y(\Omega_R\tau)$rectangular pulse phase. As the duration of the rectangular pulse increases, the population of the qubit decays exponentially. The decay rate is calculated using the
generalized Bloch equation, as described in \cite{Spin_loc, bib5, Spin_loc_multi}:

\begin{equation}
\label{eq23}
\Gamma= \frac{1}{2T_{1}} +S_{noise}(\Omega_R),
\end{equation}
where $T_{1}$ -- is the relaxation time of the qubit, $S_{noise}(\Omega_R)= (2\pi)^2/2 \int_{-\infty}^{\infty} \langle f_{01}(0) f_{01}(t)   +  f_{01}(t) f_{01}(0)  \rangle  e^{-i2 \pi \Omega_R t} dt$ is the
noise spectral density for the Rabi frequency $\Omega_R$R,
which is set by the amplitude of the driving signal, and $f_{01}$ is the transmon frequency. In this work, the energy
is measured in units of the frequency.

Thus, the spin-locking protocol allows one to
determine the noise spectral density, being easy to
implement and accurate compared to similar methods, such as Rabi spectroscopy \cite{Rabi_spec}  and generalized
spin echo \cite{CPMG}. 

The measurement results are presented in Fig.~\ref{fig:spin_lock}~(b). The data for the wet-etched sample B is shown in blue
and the data for dry-etched sample A is shown in
green. Using the experimental data, we have determined the effect of white and 1/f noise. Fig.~\ref{fig:spin_lock}(b) indicates that the amplitudes of white noise for the two
samples vary slightly. However, for dry-etched qubits,
the contribution of 1/f noise to the loss of coherence is
negligible, whereas it is significant for wet-etched
qubits. This result is in good agreement with the measurements of frequency fluctuations, which also shows
a frequency drift and strong contribution of 1/f noise
for the wet-etched sample B.

\section{CHANGINGS IN THE READOUT RESONATOR QUALITY FACTOR}

\begin{figure}[!h]
    \centering
    \includegraphics[width=0.48\textwidth]{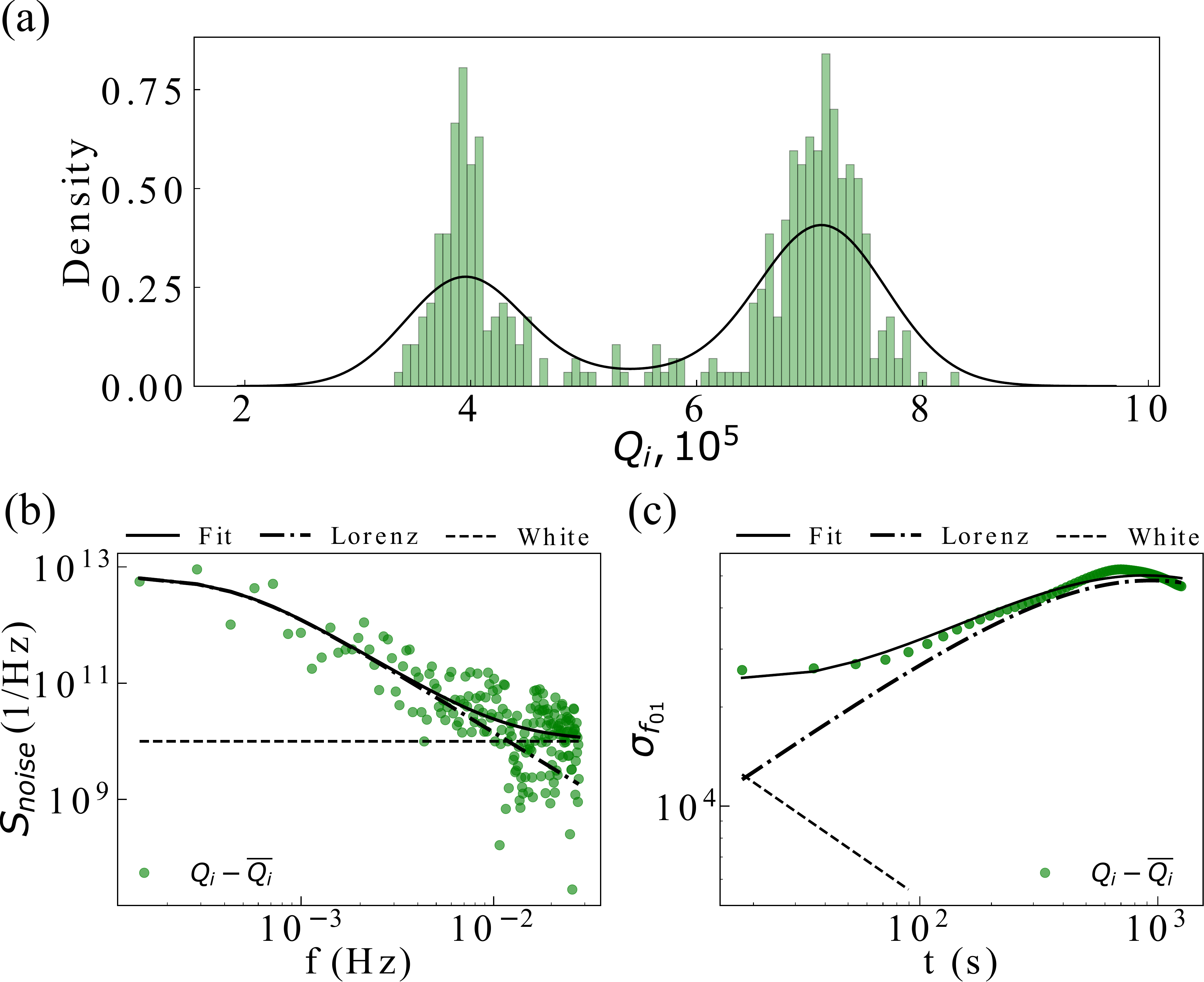}
    \caption{ (Color online) Internal quality factor measurements for one of the sample A resonators: experimental data shown in green in comparison with the numerical simulation given in black. (a) Internal quality factor histogram for the qubit in the dry-etched sample. (b) Welch spectral analysis of internal quality factor fluctuations. Different line styles correspond to the contributions from white and Lorentzian noise and their sum. (c) Allan deviation of the quality factor fluctuations; the line styles are the same as in panel (b).
    }
    \label{fig:Resonator}
\end{figure}

Two-level systems interact not only with qubits but
also with readout resonators, affecting both their frequency \cite{Burnett, Pappas} and quality factors \cite{bib8}.An example of changes in the quality factor is shown in Fig.~\ref{fig:Resonator}. Fig.~\ref{fig:Resonator} (а) presents the histogram of the time fluctuation distribution of the internal quality factor of the resonator fabricated by dry plasma chemical etching. The histogram shows two distinct peaks corresponding to the quality factor of $3.9\times10^5$ and $7.2\times10^5$, which in experiment appear as jumps in the internal quality factor of the resonator. In order to analyze the
causes of these jumps, the Welch noise spectrum and
Allan deviation were calculated; they are shown in
Figs.~\ref{fig:Resonator}(b) and (c), respectively. The experimental results can be well described by a combination of white noise and Lorentzian noise with a switching frequency of
2 mHz. Such switching frequency hints towards the
interaction of the resonator with TLS. The results of
the wet etched resonators measurements are not
shown, as no significant jumps in the Q-factor were
observed for these devices, probably because of 1/f
noise broadening of the wet-etched resonator spectral
line. Thus, individual telegraph jumps are not
resolved.

\section{RESULTS AND DISCUSSION}

In this work, the relaxation and coherence times, as
well as the fluctuations of these parameters and transmon frequencies, have been measured for two groups of transmons fabricated by dry plasma etching and wet
etching. The samples are designed the same and are
studied in an identical experimental setup. The only
difference between the two groups of transmons is the
etching method. We have found that the coherence
and relaxation times for qubits fabricated using dry
etching are significantly longer than those for qubits
fabricated using wet etching. A possible reason for
such difference is the more pronounced imperfections
of the wet etched surfaces (in particular, the sharp
edges of the superconductor electrodes and the irregularities in the aluminum film), which lead to a concentration of an electric field and a stronger coupling
to TLSs.

Analysis of the noise spectral characteristics and
Allan deviation reveals the difference in the mechanisms of dielectric losses in dry and wet etched qubits.
Studying the dry-etched qubits, we have observed
jumps in the qubit frequency and the internal quality
factor of the resonator, indicating the dominant influence of individual fluctuators. In contrast, the frequency of wet-etched qubits drifts and a significant
contribution from 1/f noise is observed in their spectra, suggesting the system being influenced by an
ensemble of fluctuators. The sharp edges formed by
wet etching not only lead to a local increase in the
electric field strength but also result in a more significant dependence of the field strength on the distance.
Consequently, TLSs located within a smaller region
contribute most. We presume that the density of defects during wet etching is higher and, consequently,
such surface cleaning process is less efficient compared to dry etching.

To summarize, the analysis of the time fluctuations
of coherence times and qubit frequencies allows one to
identify the dominant mechanisms of dielectric losses.

\section{ACKNOWLEDGMENTS}
We are grateful to A.V. Ustinov for his valuable contribution during the work and critical comments on the manuscript.

\section{FUNDING}
The design of samples and their experimental study were
supported by State Atomic Energy Corporation Rosatom
(contract no. 868-1.3-15/15-2021 dated October 5.10.2021
and contract no. 151/21-503 dated December 21, 2021,
Roadmap for Quantum Computing). The fabrication of the
samples was supported by the Ministry of Science and
Higher Education of the Russian Federation (strategic academic leadership program Priority-2030 for the National
University of Science and Technology MISIS).

\section{CONFLICT OF INTEREST}
The authors of this work declare that they have no conflicts of interest.

\section{OPEN ACCESS}
This article is licensed under a Creative Commons Attribution 4.0 International License, which permits use, sharing,
adaptation, distribution and reproduction in any medium or
format, as long as you give appropriate credit to the original
author(s) and the source, provide a link to the Creative Commons license, and indicate if changes were made. The images
or other third party material in this article are included in the
article’s Creative Commons license, unless indicated otherwise in a credit line to the material. If material is not included
in the article’s Creative Commons license and your intended
use is not permitted by statutory regulation or exceeds the
permitted use, you will need to obtain permission directly
from the copyright holder. To view a copy of this license, visit
http://creativecommons.org/licenses/by/4.0/



\small{
}
\end{document}